\input{aipcheck}

\documentclass[
    ,final            
  ]
  {aipproc}

\layoutstyle{8x11single}

\def\BBz{0$\nu\beta\beta$}

\def\mj{M{\sc ajo\-ra\-na}}
\def\dem{D{\sc e\-mon\-strat\-or}}

\def\QBB{Q$_{\beta\beta}$}

\def\ge{$^{76}$Ge}

\begin{document}

\title{Analysis techniques for background rejection at the \textsc{Majorana Demonstrator}}


\classification{23.40-s, 23.40.Bw, 14.60.Pq, 27.50.+j}
\keywords      {neutrinoless double-beta decay, germanium detector}

\newcommand{\blhill}{Department of Physics, Black Hills State University, Spearfish, SD, USA}
\newcommand{\ITEP}{Institute for Theoretical and Experimental Physics, Moscow, Russia}
\newcommand{\JINR}{Joint Institute for Nuclear Research, Dubna, Russia}
\newcommand{\lbnl}{Nuclear Science Division, Lawrence Berkeley National Laboratory, Berkeley, CA, USA}
\newcommand{\lanl}{Los Alamos National Laboratory, Los Alamos, NM, USA}
\newcommand{\uw}{Center for Experimental Nuclear Physics and Astrophysics, and Department of Physics, University of Washington, Seattle, WA, USA}
\newcommand{\unc}{Department of Physics and Astronomy, University of North Carolina, Chapel Hill, NC, USA}
\newcommand{\duke}{Department of Physics, Duke University, Durham, NC, USA}
\newcommand{\ornl}{Oak Ridge National Laboratory, Oak Ridge, TN, USA}
\newcommand{\ou}{Research Center for Nuclear Physics and Department of Physics, Osaka University, Ibaraki, Osaka, Japan}
\newcommand{\pnnl}{Pacific Northwest National Laboratory, Richland, WA, USA}
\newcommand{\ttu}{Tennessee Tech University, Cookeville, TN, USA}
\newcommand{\sdsmt}{South Dakota School of Mines and Technology, Rapid City, SD, USA}
\newcommand{\usc}{Department of Physics and Astronomy, University of South Carolina, Columbia, SC, USA}
\newcommand{\usd}{Department of Physics, University of South Dakota, Vermillion, SD, USA}
\newcommand{\ut}{Department of Physics and Astronomy, University of Tennessee, Knoxville, TN, USA}
\newcommand{\tunl}{Triangle Universities Nuclear Laboratory, Durham, NC, USA}

\author{C. Cuesta$^{a}$, N.~Abgrall$^{b}$, I.J.~Arnquist$^{c}$, F.T.~Avignone~III$^{d,e}$, C.X.~Baldenegro-Barrera$^{e}$, A.S.~Barabash$^{f}$, F.E.~Bertrand$^{e}$, A.W.~Bradley$^{b}$, V.~Brudanin$^{g}$, M.~Busch$^{h,i}$, M.~Buuck$^{a}$, D.~Byram$^{j}$, A.S.~Caldwell$^{k}$, Y-D.~Chan$^{b}$, C.D.~Christofferson$^{k}$, J.A.~Detwiler$^{a}$, Yu.~Efremenko$^{l}$, H.~Ejiri$^{m}$, S.R.~Elliott$^{o}$, A.~Galindo-Uribarri$^{e}$, T.~Gilliss$^{n,i}$, G.K.~Giovanetti$^{n,i}$, J. Goett$^{o}$, M.P.~Green$^{e}$, J.~ Gruszko$^{a}$, I.S.~Guinn$^{a}$, V.E.~Guiseppe$^{d}$, R.~Henning$^{n,i}$, E.W.~Hoppe$^{c}$, S.~Howard$^{k}$, M.A.~Howe$^{n,i}$, B.R.~Jasinski$^{j}$, K.J.~Keeter$^{p}$, M.F.~Kidd$^{q}$, S.I.~Konovalov$^{f}$, R.T.~Kouzes$^{c}$, B.D.~LaFerriere$^{c}$, J.~Leon$^{a}$, J.~MacMullin$^{n,i}$, R.D.~Martin $^{j}$, S.J.~Meijer$^{n,i}$, S.~Mertens$^{b}$, J.L.~Orrell$^{c}$, C.~O'Shaughnessy$^{n,i}$, A.W.P.~Poon$^{b}$, D.C.~Radford$^{e}$, J.~Rager$^{n,i}$, K.~Rielage$^{o}$, R.G.H.~Robertson$^{a}$, E.~Romero-Romero$^{l,e}$, B.~Shanks$^{n,i}$, M.~Shirchenko$^{g}$, N.~Snyder$^{j}$, A.M.~Suriano$^{k}$, D.~Tedeschi$^{d}$, J.E.~Trimble$^{n,i}$, R.L.~Varner$^{e}$, S. Vasilyev$^{g}$, K.~Vetter$^{r,b}$, K.~Vorren$^{n,i}$, B.R.~White$^{e}$, J.F.~Wilkerson$^{n,i,e}$, C. Wiseman$^{d}$, W.~Xu$^{o}$, E.~Yakushev$^{g}$, C.-H.~Yu$^{e}$, V.~Yumatov$^{f}$, I.~Zhitnikov$^{g}$}
{address={$^{a}$ \uw   \\
          $^{b}$ \lbnl  \\
          $^{c}$  \pnnl \\
          $^{d}$  \usc \\
          $^{e}$  \ornl \\
          $^{f}$  \ITEP \\
          $^{g}$ \JINR  \\
          $^{h}$  \duke \\
          $^{i}$  \tunl\\
          $^{j}$  \usd \\
          $^{k}$  \sdsmt \\
          $^{l}$  \ut  \\
          $^{m}$  \ou  \\
          $^{n}$ \unc \\
          $^{o}$  \lanl \\
          $^{p}$  \blhill \\
          $^{q}$  \ttu \\
          $^{r}$  Alternate address: Department of Nuclear Engineering, University of California, Berkeley, CA, USA
          }}	

\begin{abstract}
The \mj\ Collaboration is constructing the \mj\ \dem, an ultra-low background, 40-kg modular HPGe detector array to search for neutrinoless double beta decay in \ge. In view of the next generation of  tonne-scale Ge-based \BBz-decay searches that will probe the neutrino mass scale in the inverted-hierarchy region, a major goal of the \mj\ \dem\ is to demonstrate a path forward to achieving a background rate at or below 1~count/tonne/year in the 4~keV region of interest around the Q-value at 2039~keV. The background rejection techniques to be applied to the data include cuts based on data reduction, pulse shape analysis, event coincidences, and time correlations. The Point Contact design of the \dem\ $'$s germanium detectors allows for significant reduction of gamma background. 
\end{abstract}

\maketitle


\section{Introduction}
\label{sec1}

The~\mj~\dem~\cite{mjd} is an array of enriched and natural germanium detectors that will search for the neutrinoless double-beta (0$\nu\beta\beta$) decay of $^{76}$Ge. The specific goals of the~\mj~\dem~are to: Demonstrate a path forward to achieving a background rate at or below 1~cnt/(ROI-t-y) in the 4~keV region of interest (ROI) around the 2039~keV~\QBB~of the $^{76}$Ge 0$\nu\beta\beta$-decay, when scaled up to a tonne scale experiment; show technical and engineering scalability toward a tonne-scale instrument; and perform searches for other physics beyond the Standard Model, such as dark matter and axions.

The experiment is composed of 40~kg of high-purity Ge (HPGe) detectors which also act as the source of $^{76}$Ge \BBz-decay. The benefits of HPGe detectors are that Ge is an intrinsically low-background source material, understood enrichment chemistry, excellent energy resolution, and event reconstruction. P-type point contact (PPC) detectors~\cite{ppc,ppc2} were chosen after extensive R\&D by the collaboration for their powerful background rejection. 30~kg of the detectors are built from Ge material that is enriched to $>$87\% in isotope $^{76}$Ge and 10~kg are fabricated from natural Ge (7.8\% $^{76}$Ge). They enriched detector mass is approximately 850~g.

A modular instrument composed of two cryostats built from ultra-pure electroformed copper is being constructed. Each module hosts 7 strings of 3-5 detectors. The prototype module, an initial prototype cryostat fabricated from commercially produced copper, is taking data with three strings of detectors produced from natural germanium. It serves as a test bench for mechanical designs, fabrication methods, and assembly procedures that will be used for the construction of the two electroformed-copper modules. The modules are operated in a passive shield that is surrounded by a 4$\pi$ active muon veto. To mitigate the effect of cosmic rays and prevent cosmogenic activation of detectors and materials, the experiment is being deployed at 4850~ft depth (4260~m.w.e. overburden) at the Sanford Underground Research Facility in Lead, SD~\cite{surf}.

The main technical challenge of the \mj\ \dem\ is to reach a background rate of 3~cnts/(ROI-t-y) after analysis cuts, which projects to a background level of 1~cnt/(ROI-t-y) in a large scale experiment after accounting for additional improvements from thicker shielding, better self-shielding, and if necessary, increased depth. This background level represents a substantial improvement over previous generation experiments~\cite{GERDA}. To achieve this goal, background sources must be reduced and offline background rejection must be improved. The background rejection techniques to be applied to the data are described in the next sections: data reduction, pulse shape discrimination of single site events, high energy resolution, and coincident events.

\section{Run selection and data reduction}
\label{sec2}

Tools needed to perform the run evaluation and data reduction are currently being developed. First, runs are evaluated and organized in data-sets, then a data reduction framework is used to tag the different classes of events allowing the removal of non physical events.

Run selection criteria are used to create a golden \BBz-decay run list using automated checks and a further selection list is created to select detectors within these runs that are suitable for various analyses. In order to carry out an automated run selection, run selection bits that record important run information during data taking have been implemented in the acquisition. They indicate which modules are taking data, the type of run (calibration, \BBz, maintenance), if only a partial shield is in place, etc. The runs passing the run selection criteria are organized into data-sets.

A data reduction framework to tag events for its analysis efforts has been developed. The code creates a 32-bit integer for each waveform or data event. Different values are evaluated by boolean statements, and those events and waveforms that pass the filter are tagged by flipping one bit of the 32-bit integer. A complementary data reduction framework that uses neural networks is also being developed. There, Self-Organized Maps~\cite{kohonen} provide a way to classify data without any input information about the data, i.e. there is no need for training data. This method will help to identify unknown populations of events. An established list of waveform classes is in development, tag values will be refined, and their efficiencies quantified over the coming months as the first module with enriched detectors is commissioned. Also, several types of $``$external events$"$ may have an impact on the quality of Ge data. In addition to the detector data, there are $``$slow$"$ environmental events that are continuously monitored: fluctuations in the particle counts and radon levels in the experimental space, LN fills of a thermosyphon dewar, etc. These environmental data are monitored in a slow controls database, and readily accessible for a direct comparison against Ge data.

Finally, the live time measurement is the analysis process by which the number of tonne-year of exposure that the detector received is determined.

\section{Pulse shape discrimination of single site events}
\label{sec3}

The \dem\ uses PPC Ge detectors because they allow for efficient discrimination between multiple interaction sites events from gamma rays and single-site \BBz-decay events. The sharply peaked weighting potential of the point contact detector results in distinct signals from each energy deposition. The multiple interaction site events (MSE) exhibit multiple steps in the charge signal waveform and, correspondingly, multiple peaks in the current signal waveform, whereas single-site events (SSE) give a single peak in the current pulse.

\begin{figure}
 \includegraphics[width=0.45\textwidth]{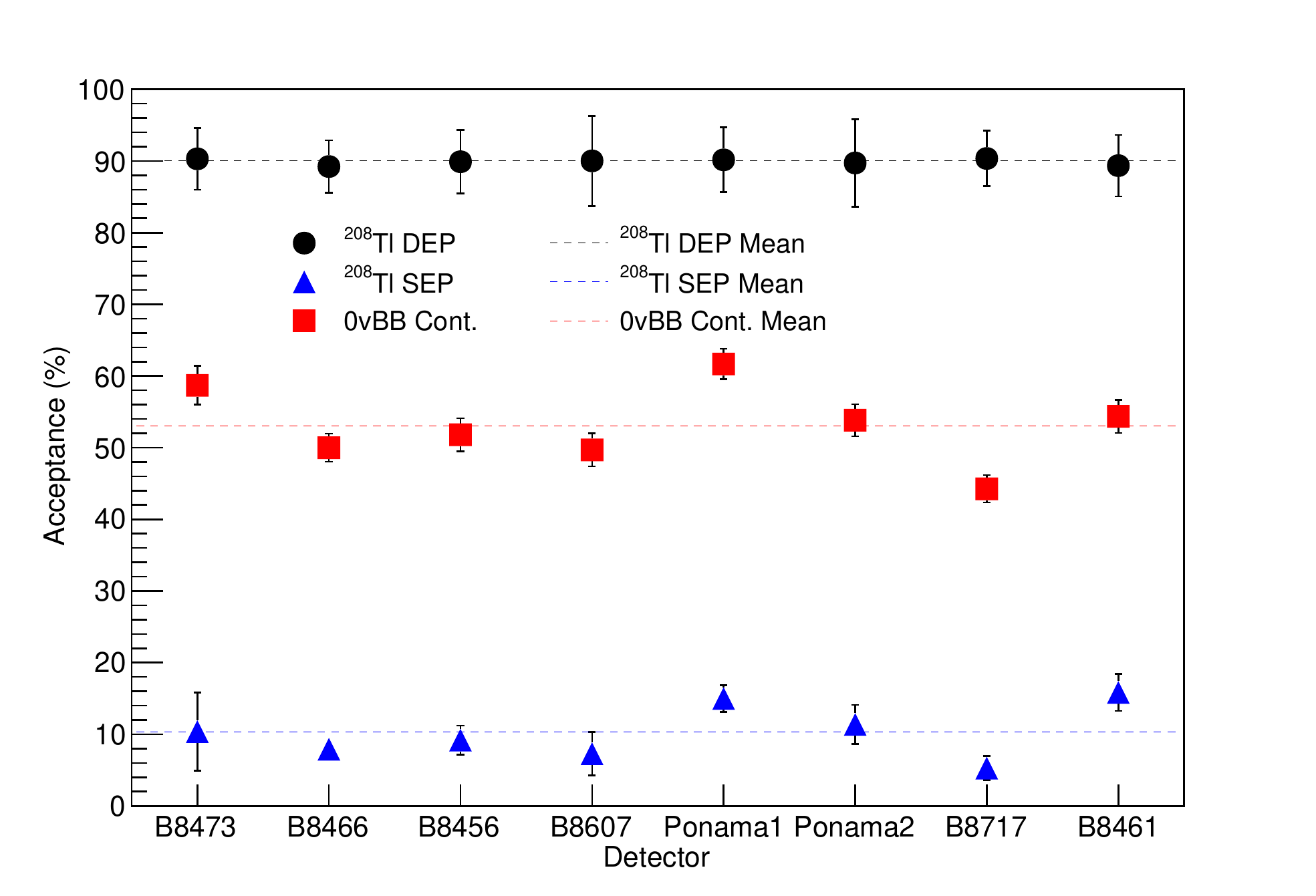}
 \includegraphics[width=0.45\textwidth]{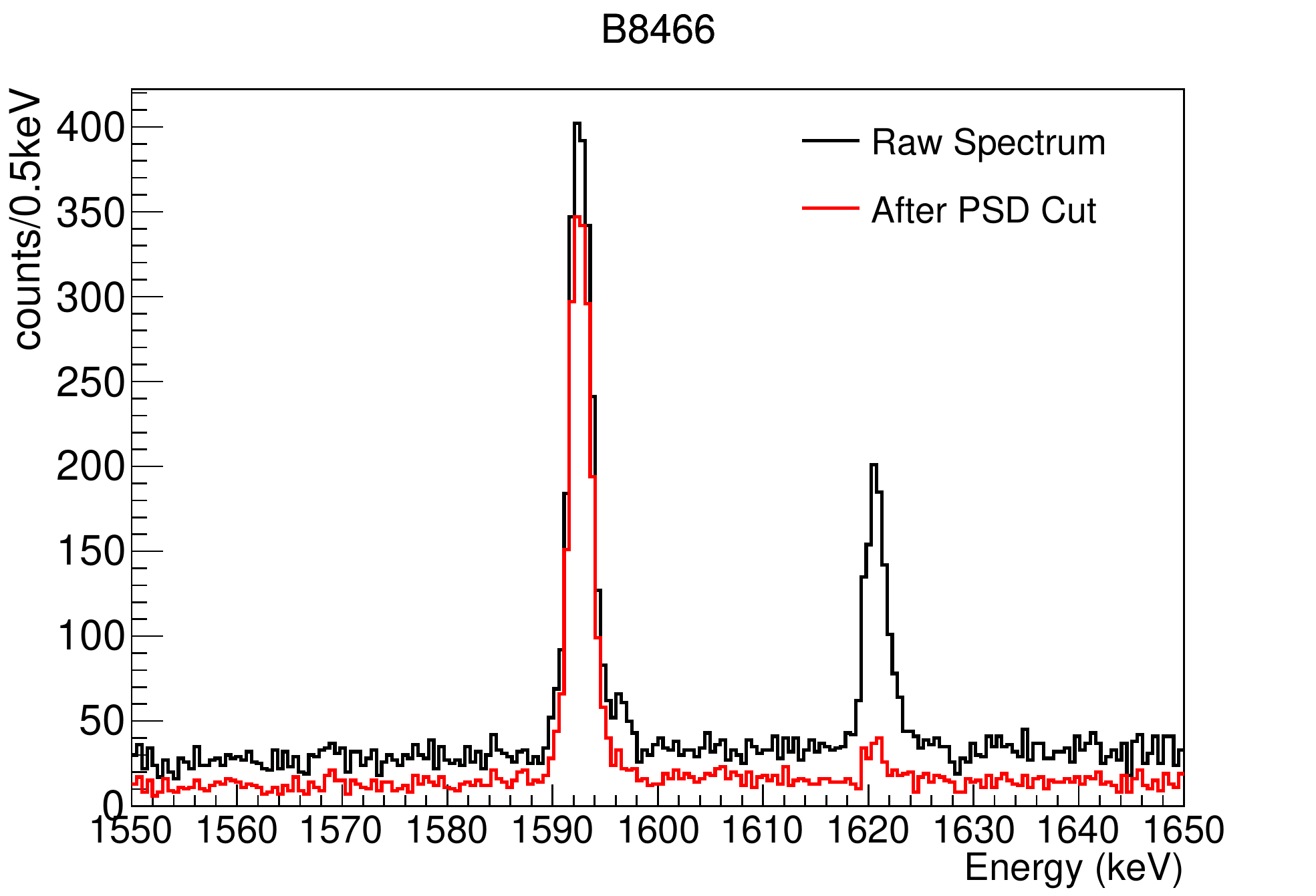}
  \caption{Left, acceptance efficiency at DEP (multi-site events), SEP (single-site events), \BBz\ ROI events of the prototype module detectors for $^{228}$Th calibration. Right, energy spectra before and after the PSD cut of B8466 Prototype Module   $^{nat}$Ge detector for $^{228}$Th calibration.}
  \label{fig:psd}
\end{figure}

Two methods to distinguish MSE from SSE using pulse shape discrimination (PSD) are considered. The first method consists of the study of the current peak amplitude to total energy (A/E)\cite{AEgerda}, i.e. the ratio of the maximum current, calculated from the time differentiation of the recorded charge waveforms, to the event total energy. With total energy deposited by multiple interactions at different locations, MSE generally have smaller maximum current for the same energy than SSE, leading to a smaller A/E. The PSD was studied and the best cut for the prototype module detectors was determined using data from a $^{228}$Th calibration. To determine the location of the cut in A/E, the acceptance efficiency of Double Escape Peak (DEP) events from 2614.5~keV gamma rays of $^{208}$Tl at 1592.5~keV was set to 90\%, and then the efficiency of the Single Escape Peak (SEP) at 2103.5~keV and in the ROI at 2039~keV were calculated. Figure~\ref{fig:psd} shows the acceptance efficiency for the prototype module detectors and the spectra before and after the PSD cut. The same analysis will be applied to the modules with enriched detectors.

A second method makes use of an algorithm that compress waveforms to a library of unique, measured single site pulse shapes. Event-by-event $\chi^{2}$ fitting of experimental pulse shapes is carried out, and depending on the fit results to the single site population, MSE and SSE are distinguished. This method performs well in preliminary tests and is in good agreement with benchmark simulations~\cite{chi2}.

To validate the pulse shape analysis techniques and estimate any associated systematic effects, pulse shape simulations are carried out with MaGe~\cite{mage}, a simulation software framework based on Geant4 developed by the \mj\ and GERDA collaborations. The resulting waveforms are unique to the detector, as it depends on the geometry and the impurities.

\section{Energy resolution}
\label{sec4}

One key advantage of HPGe detectors is their inherently excellent energy resolution ($<$0.2\% at $Q_{\beta\beta}$) associated with the low threshold for electron-pair production, leading to a narrow ROI (4~keV). Since the signal for \BBz-decay is a mono-energetic peak in the spectrum at 2039~keV, improving the resolution reduces the continuum backgrounds in the ROI. This allows for a better identification of lines in the spectrum, and minimizing the contribution from leakage of the irreducible 2$\nu\beta\beta$-decay spectrum into the ROI.

\section{Coincident events and time correlations}
\label{sec5}

\begin{figure}
 \includegraphics[width=0.6\textwidth]{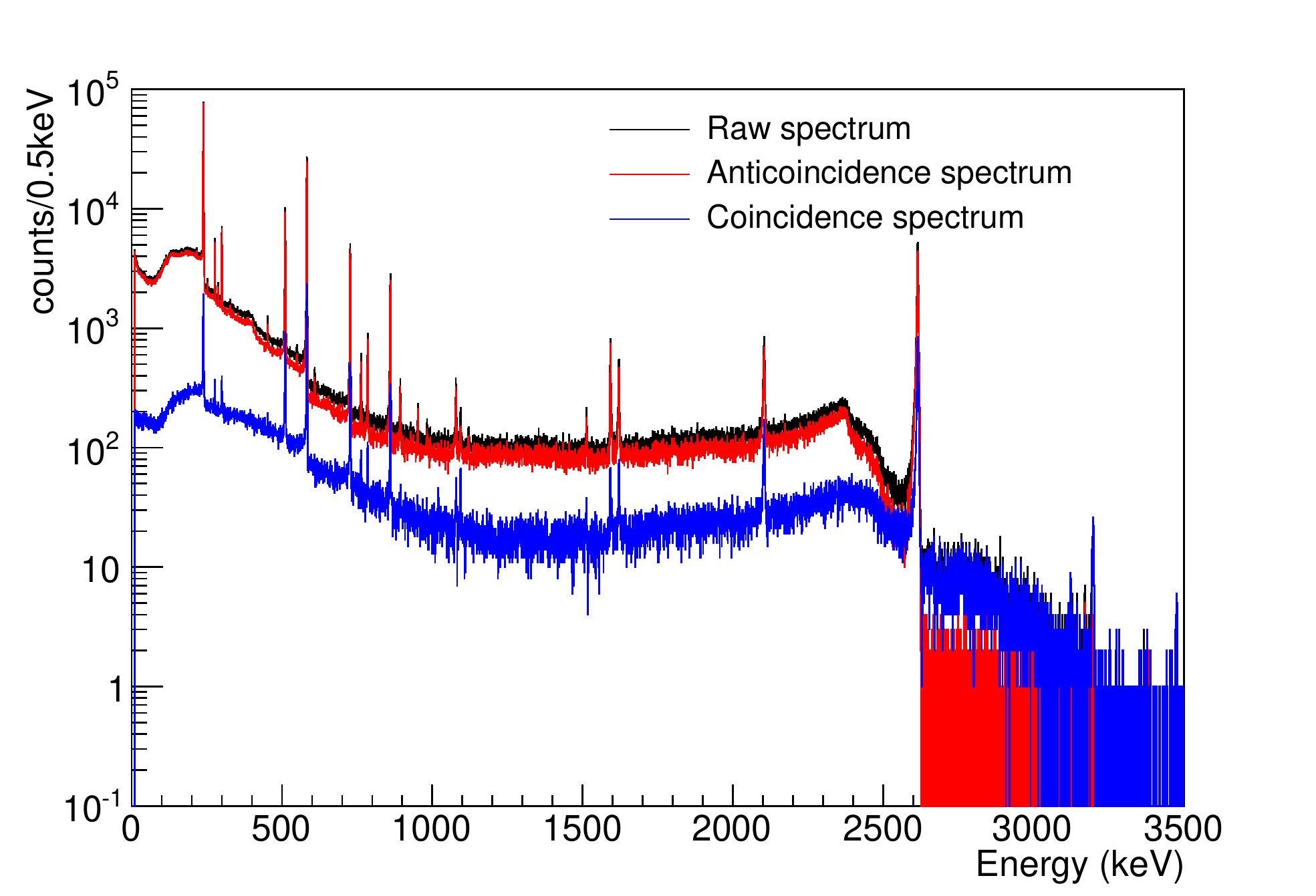}
  \caption{$^{228}$Th calibration in the prototype module: total spectrum (black), spectrum after the granularity cut (red), spectrum of multiple hit events (blue).}
  \label{fig:gran}
\end{figure}

Based on the time distribution of certain events, signatures of non plausible \BBz-decay candidates can be identified and rejected. First, events that take place simultaneously in different detectors can be rejected by the granularity cut because \BBz-decay is expected to occur locally in a single detector. Figure~\ref{fig:gran} shows this coincidence cut applied to a $^{228}$Th calibration in the prototype module and the associated reduction in background. Second, events in coincidence with a muon passing through the vetoes are also discarded. Third, the \dem\ is implementing a single-site time correlation (SSTC) cut for the $^{68}$Ge-$^{68}$Ga coincident decay for five $^{68}$Ga half-lives (T$_{1/2}$\,=\,67.71~min) following the 10~keV K-shell de-excitations. The low energy threshold plays an important role since $^{68}$Ge is a cosmogenically produced isotope in the detectors and therefore an otherwise irreducible background. The SSTC method can also be used to reduce backgrounds due to $^{208}$Tl (T$_{1/2}$\,=\,3.05~min) and $^{214}$Bi (T$_{1/2}$\,=\,19.9~min) in the Ge crystals and the inner mount.

\begin{theacknowledgments}

This material is based upon work supported by the U.S. Department of Energy, Office of Science, Office of Nuclear Physics. We acknowledge support from the Particle Astrophysics Program of the National Science Foundation. This research uses these US DOE Office of Science User Facilities: the National Energy Research Scientific Computing Center and the Oak Ridge Leadership Computing Facility. We acknowledge support from the Russian Foundation for Basic Research. We thank our hosts and colleagues at the Sanford Underground Research Facility for their support.

\end{theacknowledgments}



\bibliographystyle{aipproc}   

\bibliography{CCuesta_LRT15_mjd}

\IfFileExists{\jobname.bbl}{}
 {\typeout{}
  \typeout{******************************************}
  \typeout{** Please run "bibtex \jobname" to optain}
  \typeout{** the bibliography and then re-run LaTeX}
  \typeout{** twice to fix the references!}
  \typeout{******************************************}
  \typeout{}
 }

\end{document}